\documentclass[journal,onecolumn]{IEEEtran}
\usepackage[latin9]{inputenc}
\usepackage{amsthm}
\usepackage{amsmath}
\usepackage{cite}

\makeatletter

\theoremstyle{plain}
\newtheorem{thm}{\protect\theoremname}
\theoremstyle{definition}
\newtheorem{defn}{\protect\definitionname}
\theoremstyle{plain}
\newtheorem{lem}{\protect\lemmaname}
\theoremstyle{remark}
\newtheorem{rem}{\protect\remarkname}
\theoremstyle{plain}
\newtheorem{cor}{\protect\corollaryname}
\theoremstyle{remark}

\usepackage{amsthm,amsmath,amssymb,graphicx}
\usepackage{tikz}

\providecommand{\claimname}{Claim}
\providecommand{\corollaryname}{Corollary}
\providecommand{\definitionname}{Definition}
\providecommand{\lemmaname}{Lemma}
\providecommand{\remarkname}{Remark}
\providecommand{\theoremname}{Theorem}

\begin{document}

\title{Energy, Latency, and Reliability Tradeoffs in Coding Circuits\thanks{\noindent Part of this work was submitted for presentation at the 2016 International Symposium on Information Theory.}}
\author{Christopher G. Blake and Frank R. Kschischang\\Department of Electrical \& Computer Engineering\\University of Toronto\\ \texttt{\small christopher.blake@mail.utoronto.ca} \texttt{\small frank@comm.utoronto.ca}}
\maketitle

\begin{abstract}
It is shown that fully-parallel encoding and decoding schemes with asymptotic block
error probability that scales as $O\left(f\left(n\right)\right)$
have Thompson energy that scales as $\Omega\left(\sqrt{\ln f\left(n\right)}n\right)$.
As well, it is shown that the number of clock cycles (denoted $T\left(n\right)$)
required for any encoding or decoding scheme that reaches this bound must scale as $T\left(n\right)\ge\sqrt{\ln f\left(n\right)}$. Similar scaling results are extended to serialized computation.
 The Grover information-friction energy
model is generalized to three dimensions and the optimal energy of
encoding or decoding schemes with probability of block error $P_\mathrm{e}$
is shown to be at least $\Omega\left(n\left(\ln P_{\mathrm{e}}\left(n\right)\right)^{\frac{1}{3}}\right)$.
\end{abstract}

\section{Introduction\label{sec:Introduction}}

\IEEEPARstart {E}{xpanding} on work started in \cite{ElGamal} and more recently advanced
in \cite{groverFundamental,GroverInfoFriction,blakeConstantPePaper},
we borrow a computational complexity model introduced in \cite{Thompson}
that allows us to model the energy and number of clock cycles of a
computation. We consider fundamental tradeoffs between the asymptotic
energy, number of clock cycles, and block error probability for sequences of good encoders and decoders.

\begin{defn}
An $f\left(n\right)$\emph{-coding scheme
} is a sequence of codes of increasing block length
 $n$, together with a sequence of encoders and decoders, in which the block error
probability associated with the code of block length $n$ is less
than $f\left(n\right)$ for sufficiently large $n$.
\end{defn}

We show, in terms
of $T\left(n\right)$ (the number of clock cycles of the encoder or decoder for
the code with block length $n$) that an $f(n)$-coding scheme that
is fully parallel has encoding and decoding energy ($E$) that scales as $E\ge\Omega\left(\frac{n\ln f\left(n\right)}{T(n)}\right)$.
We show that the energy optimal number of clock cycles for encoders and decoder ($T\left(n\right)$) for an $f\left(n\right)$-coding
scheme scales as $O\left(\sqrt{\ln f\left(n\right)}\right)$, giving
a universal energy lower bound of $\Omega\left(\sqrt{\ln f\left(n\right)}n\right)$.
A special case of our result is that exponentially low probability
of error coding schemes thus have encoding and decoding energy that scales at least as
$\Omega\left(n^{\frac{3}{2}}\right)$ with energy-optimal number of clock
cycles that scales as $\Omega\left(n^{\frac{1}{2}}\right)$. This approach
is generalized to serial implementations.

Recent work on the energy complexity of good decoding has focused
largely on planar circuits. However, circuits implemented in three-dimensions
exist \cite{3DCircuitBook}, and so we generalize the recent
information friction (or bit-meters) model introduced by Grover in
\cite{GroverInfoFriction} to circuits implemented in three-dimensions
and extend the technique of Grover to show that, in terms of block length $n$, a bit-meters coding
scheme in which block error probability is given by $P_{\mathrm{e}}(n)$
has encoding/decoding energy that scales as $\Omega\left(n\left(\ln P_{\mathrm{e}}\left(n\right)\right)^{\frac{1}{3}}\right)$.
We show how this approach can be generalized to an arbitrary number
of dimensions.

 In Section \ref{sec:Computational-Complexity-Lower}
 we discuss prior work, and in particular
 we discuss existing results on complexity lower bounds for different
 models of computation for different notions of ``good'' encoders
 and decoders. The main technical results of this work are in Section \ref{sec:ThompsonModelSection},
where we study the Thompson energy model, and in Section \ref{sec:Information-Friction-Section},
where we study a multi-dimensional generalization of the Grover bit-meters
model. In these sections we present lower bounds for decoders, as the derivation for encoding lower bounds is almost exactly the same. We provide an outline of the technique for encoder lower bounds in Section~\ref{sec:encoderLowerBounds}. In Section~\ref{sec:Limitations-of-Result} we discuss limitations
and weaknesses in the model used. In Section~\ref{sec:Other-Energy-Models}, we discuss
other energy models of computation. In Section~\ref{sec:Future-Work}
we discuss possible future work, and conjecture that similar tradeoffs
may extend to circuits that perform inference. 

\emph{Notation:} We use standard Bachmann-Landeau
notation in this paper. The statement $f(x)=O(g(x))$
means that for sufficiently large $x$, $f(x)\le cg(x)$ for some
positive constant $c$. The statement $f(x)=\Omega(g(x))$ means that
for sufficiently large $x$, $f(x)\ge cg(x)$ again for some constant
$c$. The statement $f(x)=\Theta(g(x))$ means that there are two
positive constants $b$ and $c$ such that $b\le c$ and for sufficiently
large $x$, $bg(x)\le f(x)\le cg(x)$.

\section{Prior Related Work: Computational Complexity Lower Bounds for Good Decoders and Encoders\label{sec:Computational-Complexity-Lower}}

The earliest work on computational complexity lower bounds for good decoding comes from Savage in \cite{savage1969transactionsArticle}
and \cite{savagePart2}, which considered bounds on the memory requirements
and number of logical operations needed to compute decoding functions.
However, wiring area is a fundamental cost of good decoding and the
authors do not consider this. More recently, in \cite{ElGamal}, the
authors use a model similar to our model, except the notion of ``area''
the authors use is the size of the smallest rectangle that completely
encloses the circuit under consideration. 

In \cite{groverFundamental}, Grover \emph{et al.} consider the same model that
we do, and find Thompson energy lower bounds as a function of probability
of block error probability for good encoders and decoders. Our analysis of the
Thompson model differs from the approach of Grover \emph{et al. }in
a number of ways. Firstly, central to the work of Grover \emph{et
	al.} is a bound on block error probability if inter-subcircuit bits
communicated is low (presented in Lemma 2 in the Grover \emph{et al.}
paper), which is analogous to our result in (\ref{eq:theMainArgument})
of the proof of Theorem~\ref{thm:TheoremRelatingfofntoEnergy}. Our
result simplifies this relationship using simple probability arguments.
Secondly, the Grover \emph{et al.} paper does not present what energy-optimal
number of clock cycles are in terms of asymptotic probability of block
error, nor do they present the fundamental tradeoff between number of clock cycles,
energy, and reliability within the Thompson model that we present
in this paper. Moreover, the technique of \cite{groverFundamental} does not extend to serial implementations.

In \cite{blakeConstantPePaper} we considered the corner case of decoding schemes in
which block error probability asymptotically was less than $\frac{1}{2}$ for serial and parallel decoding schemes.  We did not, however, analyze schemes in terms of the rate at which block error probability approaches $0$, nor did we compute energy-optimal number of clock cycles as we do herein.

There has also been some work on complexity scaling rules for encoding and decoding of specific types of codes. Low density parity check coding VLSI scaling rules have been studied in \cite{BlakeLDPCAlmostSure, ganesanPaper} and polar coding scaling rules have been studied in \cite{blakeKschischangPolarScaling}. The scaling rules presented in this paper are general and apply to any code.

Another computational model that has proven more tractable than the
Turing Time complexity model is the constant depth circuit model (see
\cite{AroraBarakComputationalComplexityBook} for a detailed description
of this model).  Super-polynomial lower
bounds on the size of constant depth circuits that compute certain
notions of ``good encoding functions'' (though not decoding) were
derived in \cite{lovettViolaCannotSampleGoodCodes}. In this case,
the notion of ``good'' considered was the ability to correct at
least $\Omega\left(n\right)$ errors at rates asymptotically above
$0$. Similar related work exists in \cite{rychkov} which discovered
lower bounds on the formula-size of functions that perform good error
control coding; similar bounds were later discovered in \cite{kojevnikov}.

\section{Thompson Model\label{sec:ThompsonModelSection}}

\subsection{Circuit Model\label{sub:Circuit-Model}}

The model we will consider derives from Thompson \cite{Thompson}. The specific model we consider has been studied in  \cite{blakeConstantPePaper,groverFundamental,BlakeLDPCAlmostSure,ganesanPaper}. The reader should refer to \cite{blakeConstantPePaper} for details of the model. The important parameters to be extracted from the model are $A$, the circuit area, and $T$, the number of clock cycles in a computation. Since in this paper we are only concerned with scaling rules, we assume that both the technology constant and the wire width considered in \cite{groverFundamental,blakeConstantPePaper} are equal to $1$. The energy of a computation is thus defined as $E=AT$.

\begin{figure} \centering 

\includegraphics[width= 3 in]{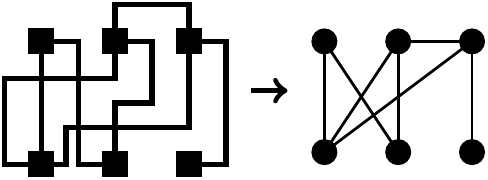}

\caption{A circuit next to its associated graph.
}
\label{fig:circuitIsAGraphDiagram}  \end{figure}

Note that a circuit can be associated with a graph in the natural
way, in which a wire corresponds to an edge of the graph and a node
corresponds to a vertex. An edge connects two vertices if their associated
nodes are connected by wires. A diagram of a small circuit next to
its associated graph is given in Fig. \ref{fig:circuitIsAGraphDiagram}.

Lemma~\ref{lem:ThompsonLemma} presented below is derived in  \cite{Thompson} and it relates the area of a circuit to its graph's minumum bisection width, and is a key component of our Thompson model circuit lower bounds.

\subsection{Definitions and Lemmas}

To present the main results of this paper we shall present a sequence
of definitions and lemmas similar to \cite{blakeConstantPePaper,groverFundamental}.

\begin{lem}
\label{lem:GuessingLemma} \cite{blakeConstantPePaper} Suppose that $X$, $Y$, and $\hat{X}$
are random variables that form a Markov chain $X\rightarrow Y\rightarrow\hat{X}$
and $X$ takes on values from a finite alphabet $\mathcal{X}$ with
a uniform distribution, (i.e., $P\left(X=x\right)=\frac{1}{\left|\mathcal{X}\right|}$
for all $x\in\mathcal{X}$), $Y$ takes on values from a finite set
$\mathcal{Y}$, and $\hat{X}$ from a set $\mathcal{\hat{X}}$. Suppose
as well that $\mathcal{\hat{X}}\in\mathcal{X}$. Then:

\[
P\left(\hat{X}=X\right)\le\frac{\left|\mathcal{Y}\right|}{\left|\mathcal{X}\right|}.
\]
\end{lem}
\begin{rem}
We will interpret $X$ as the set of symbols a particular subcircuit
will need to estimate, $\hat{X}$ as that subcircuit's estimate of
those symbols, and $Y$ as the bits injected into the subcircuit during
the computation. Note that this result mirrors the result of Lemma
4 in \cite{GroverInfoFriction}. In this lemma, the author proves
that if a circuit has $\frac{r}{3}$ bits to make an estimate $\hat{X}$
of a random variable $X$ that is uniformly distributed over all binary
strings of length $r$, then that circuit makes an error with probability
at least $\frac{1}{9}$. Our lemma presented here 
includes this lemma as a special case by setting $\left|\mathcal{Y}\right|=2^{\frac{r}{3}}$
and $\left|\mathcal{X}\right|=2^{r}$. In this case we can infer:
$P\left(\hat{X}\ne X\right)\ge1-\frac{2^{\frac{r}{3}}}{2^{r}}\ge1-2^{-\frac{2}{3}r}>\frac{1}{9}$,
where the last inequality is implied by $r\ge1$.\end{rem}
\begin{IEEEproof}
(of Lemma \ref{lem:GuessingLemma}) See \cite{blakeConstantPePaper}.
This flows from a simple application of the law of total probability
and the definition of a Markov chain. \end{IEEEproof}
\begin{defn}
A\emph{ bisection} of a graph $G=\left(V, E\right)$ of a set of vertices
$V'\in V$ is a set of edges $E'\in E$ that, once removed from the
graph, results in two disconnected subgraphs with vertices $V_{1}$
and $V_{2}$ in which $\left|\left|V'\cap V_{1}\right|-\left|V'\cap V_{2}\right|\right|\le1$.
That is, it is the set of edges that, once removed, divides the vertices
of $V'$ roughly in half. The \emph{minimum bisection width} of a
set of vertices $V'$ is the size of a smallest bisection.
\end{defn}
Note that since a circuit is associated with a graph, we can discuss
such a circuit's minimum bisection width, that is the minimum bisection
width of the graph with which it is associated. Herein we will consider
bisecting the output nodes of a circuit.

\begin{lem}
\label{lem:ThompsonLemma}All circuits whose associated graphs have
minimum bisection width $\omega$ have circuit area $A\ge\frac{\omega^{2}}{4}$.\end{lem}
\begin{IEEEproof}
See Thompson \cite{Thompson}.
\end{IEEEproof}

We now discuss the notion of nested minimum bisection, a concept introduced by Grover \emph{et al.} in \cite{groverFundamental} and also used in \cite{blakeConstantPePaper} which we again present here so the paper is self contained.

Suppose that a circuit has $k$ output nodes. If the output nodes
of such a circuit are minimum bisected, this results in two disconnected
subcircuits each with, roughly, $\frac{k}{2}$ output nodes. These
two subcircuits can each have their output nodes minimum bisected
again, resulting in four disconnected subcircuits, now each with roughly
$\frac{k}{4}$ output nodes. 
\begin{defn}
This process of nested minimum bisections on a circuit, when repeated $r$ times, is called \emph{performing
$r$-stages of nested minimum bisections.} In the case of this paper,
the set of nodes to be minimum bisected will be the output nodes.
We may also refer to this process as \emph{performing nested bisections,
}and a circuit under consideration in which nested bisections have
been performed as a \emph{nested bisected circuit}. Note that we will
omit the term ``minimum'' in discussions of such objects, as this
is implicit.
\end{defn}
Note that associated with an $r$-stage nested bisected circuit are
$2^{r}$ subcircuits. Note as well that once a subcircuit has only
one node, it does not make sense to bisect that subcircuit again.
Suppose we are nested-bisecting the $k$ output nodes of a circuit.
In this case, one cannot meaningfully nested-bisect the output nodes
of a circuit $r$ times if $2^{r}>k$.

Note that each of the $2^{r}$ subcircuits induced by the $r$-stage
nested bisection may have some internal wires, and also wires that
were deleted and connect to nodes in other subcircuits. We can index
the $2^{r}$ subcircuits with the symbol $i$.
\begin{defn}
Let the number of wires attached to nodes in subcircuit $i$ that
were deleted in the nested bisections be $f_{i}$. This quantity is
the \emph{fan-out of subcircuit $i$.}
\end{defn}
We shall also consider the bits communicated to a given subcircuit.
\begin{defn}
\label{def:biBitsCommunicatedToiThSubcircuit}Let $b_{i}=f_{i}T$,
where we recall that $T$ is the number of clock cycles used in the
running of the circuit under consideration. This quantity is called
the \emph{bits communicated to the $i$th subcircuit.}
\end{defn}
We can now define an important quantity.
\begin{defn}
The quantity $B_{r}=\sum_{i=1}^{2^{r}}b_{i}$ is the \emph{inter-subcircuit
bits communicated. }
\end{defn}
Note that each subcircuit induced by the nested bisections will each
have close to $\frac{k}{2^{r}}$ output nodes within them (a consequence
of choosing to bisect the output nodes at each stage), however, each
may have a different number of input nodes.
\begin{defn}
This quantity is called the \emph{number of input nodes in the $i$th
subcircuit} and we denote it $n_{i}$. 
\end{defn}
Note that $\sum_{i=1}^{2^{r}}n_{i}=n$ for all valid choices of $r$.
That is, the sum over the number of input nodes in each subcircuit
is the total number of input nodes in the original circuit.

This now allows us to present an important lemma.
\begin{lem}
\label{lem:GroverATsquaredLemma}All fully-parallel circuits with
inter-subcircuit bits communicated $B_{r}$ have product $AT^{2}$
bounded by:

\begin{equation}
AT^{2}\ge\frac{\left(\sqrt{2}-1\right)^{2}}{32}\frac{B_{r}^{2}}{2^{r}}=c_{1}\frac{B_{r}^{2}}{2^{r}}\label{eq:ATsquaredBound}
\end{equation}
where we define $c_{1}=\frac{\left(\sqrt{2}-1\right)^{2}}{32}$.\end{lem}
\begin{IEEEproof}
This result, from Grover \emph{et al.} \cite{groverFundamental} flows from applying
Lemma \ref{lem:ThompsonLemma} recursively on the nested-bisected
structure and optimizing. \end{IEEEproof}
\begin{lem}
\label{lem:energyBrLemma}All fully-parallel circuits with inter-subcircuit
bits communicated $B_{r}$ and number of input nodes $n$ have product
$AT$ bounded by:
\[
AT\ge c_{2}\sqrt{\frac{n}{2^{r}}}B_{r}
\]
where we define $c_{2}=\frac{\sqrt{2}-1}{4\sqrt{2}}$.\end{lem}
\begin{IEEEproof}
See \cite{groverFundamental}. This result flows from the
observation that $A\ge n$ for a fully parallel circuit and then combining
this inequality with (\ref{eq:ATsquaredBound}).\end{IEEEproof}
\begin{defn}
An \emph{$(n,k)$-decoder }is a circuit that computes a decoding function
$f:\left\{ 0,1\right\} ^{n}\rightarrow\left\{ 0,1\right\} ^{k}$.
It is associated with a codebook, (and therefore, naturally, an encoding
function, which computes a function $g:\left\{ 0,1\right\} ^{k}\rightarrow\left\{ 0,1\right\} ^{n}$),
a channel statistic, $P\left(y^{n}|x^{n}\right)$ (which we will assume
herein to be the statistic induced by $n$ channel uses of a binary
erasure channel), and a statistic from which the source is drawn $p\left(x^{k}\right)$
(which we will assume to be the statistic generated by $k$ independent
fair binary coin flips). The quantity $n$ is the block length of
the code, and the quantity $k$ is the the number of bits decoded. 
\end{defn}
\begin{defn}
The \emph{block error probability} of a decoder, denoted $P_{\mathrm{e}}$,
is the probability that the decoder's estimate of the original source
is incorrect. Note that this probability depends on the source distribution,
the channel, and the function that the decoder computes. 
\end{defn}
\begin{defn}
A \emph{decoding scheme} is an infinite sequence of circuits $D_{1},D_{2,}\ldots$
each of which computes a decoding function, with block lengths $n_{1}<n_{2}<\ldots$
and bits decoded $k\left(n_{1}\right),k\left(n_{2}\right),\ldots$.
They are associated with a sequence of codebooks $C_{1},C_{2},\ldots$
and a channel statistic. 
\end{defn}
We assume throughout this paper that the channel statistic associated
with each decoder is the statistic induced by $n$ uses of a binary
erasure channel. Our lower bound results also apply to any channel
that is a degraded erasure channel, including the binary symmetric
channel. Our results in terms of binary erasure probability $\epsilon$
can be applied to decoding schemes for the binary symmetric channel
with crossover probability $p$ by substituting $p=2\epsilon$.
\begin{defn}
We let $P_{e}\left(n\right)$ denote the\emph{ block error probability
}for the decoder with input size $n$. We let $R\left(n\right)=\frac{k\left(n\right)}{n}$
be the \emph{rate} of the decoder with input size $n$.
\end{defn}
We also classify decoding schemes in terms of how their probability
of error scales in the definition below.
\begin{defn}
An $f\left(n\right)$\emph{-decoding scheme }is a decoding scheme
in which for sufficiently large $n$ the block error probability $P_{\mathrm{e}}(n)< f(n)$.
\end{defn}
\begin{defn}
The \emph{asymptotic-rate, }or more compactly, the \emph{rate }of
a decoding scheme is $\lim_{n\rightarrow\infty}R\left(n\right)$, if this limit exists, which
we denote $R$. 
\end{defn}
Note that the rate of a decoding scheme may not be the rate of any
particular codebook in the decoding scheme.
\begin{defn}
An \emph{exponentially-low-error decoding scheme} is an $e^{-cn}$-decoding scheme for some $c>0$ with asymptotic rate $R$ greater
than $0$.
\end{defn}
We will also consider another class of decoding schemes, one which
can be considered less reliable.
\begin{defn}
A \emph{polynomially-low-error decoding scheme} is a $\frac{1}{n^{t}}$-decoding
scheme for some $t>0$ with asymptotic rate $R>0$.
\end{defn}
We will also need to define a sublinear function, which will be used
to deal with a technicality in Theorem~\ref{thm:TheoremRelatingfofntoEnergy}.
\begin{defn}
A sublinear function $f\left(n\right)$ is a function in which $\lim_{n\rightarrow\infty}\frac{f\left(n\right)}{n}=0$.

\subsection{Main Lower Bound Results}
We can now state the main theorem of this paper.\end{defn}
\begin{thm}
\label{thm:TheoremRelatingfofntoEnergy}All $f\left(n\right)$-decoding
schemes associated with a binary erasure channel with erasure probability
$\epsilon$ in which $f\left(n\right)$ monotonically decreases to
$0$ and in which $-\ln\left(f\left(n\right)\right)$ is a sublinear
function have energy that scales as 
\begin{equation}
E\ge c_{3}\sqrt{\frac{\ln\left(f\left(n\right)\right)}{\ln(\epsilon)}}k\label{eq:energyLowerBound}
\end{equation}
where $c_{3}=\frac{\sqrt{\ln2}(\sqrt{2}-1)}{16\sqrt{2}}$ and $AT^{2}$ complexity
that scales as:

\begin{equation}
AT^{2}\ge c_{4}\frac{k^{2}\ln\left(f\left(n\right)\right)}{n\ln(\epsilon)}\label{eq:ATsquaredTheoremAsFunctionOfFn}
\end{equation}
for another positive constant $c_{4}=\frac{\ln(2)\left(\sqrt{2}-1\right)^{2}}{512}$.\end{thm}
\begin{IEEEproof}
Associated with each decoder is its $B_{r}$, the inter-subcircuit
bits communicated. We can choose $r$ to be any function of $n$ so
long as $2^{r}<nR\left(n\right)=k\left(n\right)$. From here on, we
will suppress the dependence of $r\left(n\right)$, $k\left(n\right)$,
and $R\left(n\right)$ on $n$. For ease
of notation, let $N=2^{r}$ be the number of subcircuits induced by
the $r$-stages of nested bisections. Consider any specific sufficiently
large circuit in our decoding scheme, and suppose that $B_{r}<\frac{k}{2}$.
Then there exists at least $\frac{N}{2}$ subcircuits in which $b_{i}<\frac{k}{N}$
(where we recall $b_{i}$ is the bits communicated to the $i$th subcircuit
from Definition~\ref{def:biBitsCommunicatedToiThSubcircuit}). Suppose
not, \emph{i.e.}, that there are $\ge\frac{N}{2}$ subcircuits with
$b_{i}\ge\frac{k}{N}$. Then, $B_{r}\ge\frac{k}{N}\frac{N}{2}=\frac{k}{2}$,
violating the assumption that $B_{r}<\frac{k}{2}$. Call the set of
at least $\frac{N}{2}$ subcircuits with bits communicated to them
less than $\frac{k}{N}$ $Q$. Using a similar averaging argument, we claim that within
$Q$ there must be one subcircuit in which $n_{i}\le\frac{2n}{N}$.
If not, if all $\frac{N}{2}$ subcircuits in $Q$ have greater than
$\frac{2n}{N}$ input bits injected into them, then the total number
of inputs nodes in the entire circuit is greater than $\frac{2n}{N}\frac{N}{2}=n$,
but there are only $n$ input nodes in the entire circuit. Thus, there
is at least one subcircuit in $Q$ in which $b_{i}<\frac{k}{N}$ and
$n_{i}\le\frac{2n}{N}$.

Suppose that all the input bits injected into this special subcircuit
are erased. Then, that subcircuit makes an error with probability
at least $\frac{1}{2}$ by Lemma~\ref{lem:GuessingLemma}, since it
will have to form an estimate of $\frac{k}{N}$ bits by only having
injected into it fewer than $\frac{k}{N}$ bits. Thus, if $B_{r}\le\frac{k}{2}$
then:
\begin{eqnarray*}
P_{e} & \ge & P\left(\mathrm{error|\mathrm{all\ n_{i}\ bits\ erased}}\right)P\left(\mathrm{all\ n_{i}\ bits\ erased}\right)\\
 & \ge & \frac{1}{2}\epsilon^{n_{i}}
\end{eqnarray*}
where this first inequality flows from summing one term in a law of
total probability expansion of the probability of block error, and
the second from lower bounds on these probabilities.

Combining this observation with the fact the $n_{i}\le\frac{2n}{N}$
gives us the following observation:
\begin{equation}
\mathrm{if\ }B_{r}\le\frac{k}{2}\mathrm{\ then\ }P_{\mathrm{e}}\ge\frac{1}{2}\epsilon^{n_{i}}\ge\frac{1}{2}\epsilon^{\frac{2n}{N}}\label{eq:theMainArgument}
\end{equation}
This is true for any valid choice of $r$.

Now suppose that our decoding scheme is an $f\left(n\right)$-decoding
scheme.
We choose $r$ to be
\[
r=\left\lfloor \log_{2}\frac{2n\ln(\epsilon)}{\ln(2)\ln\left(f\left(n\right)\right)}\right\rfloor 
\]
so that 
\begin{equation}
N=2^{r}\approx\frac{2n\ln(\epsilon)}{\ln (2)\ln\left(f\left(n\right)\right)}.\label{eq:divisionIntoN}
\end{equation}
This is a valid choice of $r$ because $N$ cannot grow faster than
$O\left(n\right)$ because we assumed $P_{e}\left(n\right)$ was monotonically
decreasing (easily checked by inspection). Note as well that $N$
increases with $n$ because of the sub-linearity assumption of $-\ln\left(f(n)\right)$.
Then, if $B_{r}\le\frac{k}{2}$, by directly substituting into (\ref{eq:theMainArgument}),
\begin{eqnarray*}
P_{e} & \ge & \frac{1}{2}\exp\left(\frac{\ln(\epsilon)2\ln(2) n\ln\left(f(n)\right)}{2n\ln(\epsilon)}\right)\\
 & = & \frac{1}{2}\exp\left(\ln(2)\ln\left(f(n)\right)\right)=f\left(n\right).
\end{eqnarray*}
In other words, if $B_{r}\le\frac{k}{2}$ then our decoding scheme
is not an $f(n)$-decoding scheme. Thus, for this choice of $r$,
$B_{r}>\frac{k}{2}$. 

Thus, by Lemma~\ref{lem:energyBrLemma},
\begin{eqnarray*}
E & \ge & c_{2}\sqrt{\frac{n}{2^{\left\lfloor \log_{2}\frac{2n\ln(\epsilon)}{\ln(2)\ln\left(f(n)\right)}\right\rfloor }}}\frac{k}{2}\\
 & \ge & c_{2}\sqrt{\frac{n}{2^{\log_{2}\frac{2n\ln(\epsilon)}{\ln(2)\ln\left(f(n)\right)}+1}}}\frac{k}{2}\\
 & \ge & c_{2}\sqrt{\frac{n}{2\left(\frac{2n\ln(\epsilon)}{\ln(2)\ln\left(f(n)\right)}\right)}}\frac{k}{2}\\
 & \ge & c_{3}\sqrt{\frac{\ln\left(f(n)\right)}{\ln(\epsilon)}}k\\
\end{eqnarray*}
where we substituted the value for $N$ in the first line, used the
fact that $\left\lfloor x\right\rfloor \le x+1$ in the second, and
simplified the lines that followed, proving inequality (\ref{eq:energyLowerBound})
of the theorem. As well, by Lemma~\ref{eq:ATsquaredBound}, using
$B_{r}>\frac{k}{2}$ for this choice of $r$, following a similar
substitution as in the previous paragraph: 
\begin{eqnarray*}
AT^{2} & \ge & c_{1}\frac{B_{r}^{2}}{2^{r}}.\\
 & \ge & c_{1}\frac{k^{2}}{4\left(2^{\left\lfloor \log_{2}\frac{2n\ln(\epsilon)}{\ln(2)\ln\left(f(n)\right)}\right\rfloor }\right)}\\
 & \ge & c_{1}\frac{k^{2}}{4\left(2^{\log_{2}\frac{2n\ln(\epsilon)}{\ln(2)\ln\left(f(n)\right)}+1}\right)}\\
 & = & c_{1}\frac{k^{2}}{8\left(\frac{2n\ln(\epsilon)}{\ln(2)\ln\left(f(n)\right)}\right)}\\
 & = & \frac{c_{1}\ln(2)}{16}\frac{k^{2}\ln\left(f(n)\right)}{\ln(\epsilon)}
\end{eqnarray*}
and the inequality in (\ref{eq:ATsquaredTheoremAsFunctionOfFn}) flows
from substituting the appropriate value for $c_{1}$ as defined in
Lemma~\ref{eq:ATsquaredBound}.
\end{IEEEproof}
\begin{cor}
\label{cor:exponentialCorollary}All exponentially low error decoding
schemes have energy that scales as 
\[
E\ge\Omega\left(\frac{n^{\frac{3}{2}}}{p\left(n\right)}\right)
\]
 for all functions $p\left(n\right)$ that increase without bound.
In other words, all exponential probability of error decoding schemes
have energy at least that scales very close to $\Omega\left(n^{\frac{3}{2}}\right)$.
Moreover, any such scheme that has energy that grows optimally, i.e.
as $AT=O\left(n^{\frac{3}{2}}\right)$, must have $T\left(n\right)\ge\Omega\left(n^{0.5}\right)$. \end{cor}
\begin{IEEEproof}
Note that an exponentially low error decoding scheme has $P_{e}\le e^{-cn}$.
Thus, such a scheme is also an $e^{-c\frac{n}{p\left(n\right)}}$-decoding
scheme, for any increasing $p\left(n\right)$. The result then directly
flows by substituting $f\left(n\right)=e^{-c\frac{n}{p\left(n\right)}}$
into (\ref{eq:energyLowerBound}) of Theorem~\ref{thm:TheoremRelatingfofntoEnergy}.

For the second part of the corollary, suppose that for some constant
$c$, a decoding scheme has
\begin{equation}
AT=\Theta (n^{\frac{3}{2}} ).\label{eq:ATBoundforExponential}
\end{equation}
We have as well from (\ref{eq:ATsquaredTheoremAsFunctionOfFn}) and
substituting $f(n)=e^{-c\frac{n}{p\left(n\right)}}$

\begin{equation}
AT^{2}\ge \Omega \left( \frac{n^{2}}{p\left(n\right)} \right) \label{eq:ATsquaredForExponentialBound}
\end{equation}
where we use the fact that $k=Rn$ (since
by definition exponentially-low error decoding schemes have asymptotic
rate greater than $0$).

Suppose that 
\begin{equation}
T< O\left( \frac{n^{\frac{1}{2}}}{g\left(n\right)} \right) \label{eq:Tbound}
\end{equation}
for a $g\left(n\right)$ that grows with $n$, \emph{i.e.,} that $T$
asymptotically grows slower than $O\left(n^{\frac{1}{2}}\right)$. Then, to
satisfy (\ref{eq:ATsquaredForExponentialBound}) we need

\begin{equation}
AT^{2}\ge \Omega \left( \frac{n^{2}}{p\left(n\right)} \right)\label{eq:toBeUnsatisfied}
\end{equation}
for all increasing $p\left(n\right)$, implying 
\[
A\ge \Omega \left( \frac{ng\left(n\right)^{2}}{p\left(n\right)} \right).
\]
To see this precisely, suppose otherwise and then it is easy to see
that, combined with (\ref{eq:Tbound}) the inequality in (\ref{eq:toBeUnsatisfied})
will be unsatisfied. If this is true, however, then the product 
\[
AT\ge \Omega \left( \frac{ng\left(n\right)^{2}}{p\left(n\right)}\frac{n^{\frac{1}{2}}}{g\left(n\right)} \right) = \Omega \left( \frac{n^{\frac{3}{2}}g\left(n\right)}{p\left(n\right)} \right).
\]

Since this is true for all increasing $p\left(n\right)$, it is true
for, say, $p\left(n\right)=\ln g\left(n\right)$, implying that the
product $AT$ grows strictly faster than $\Omega\left(n^{\frac{3}{2}}\right)$,
contradicting the assumption of (\ref{eq:ATBoundforExponential}).
\end{IEEEproof}

We generalize Corollary~\ref{cor:exponentialCorollary} to decoding schemes with different asymptotic block error probabilities below:
\begin{thm}\label{thm:generalTheoremForoptimalT}
All $f(n)$-decoding schemes with asymptotic rate greater than $0$ in which $f(n)$ is sub-exponential with energy that scales as $E=\Theta\left(\sqrt{\ln f(n)} n\right)$ (that is, their energy matches the lower bound of (\ref{eq:energyLowerBound}) of Theorem~\ref{thm:TheoremRelatingfofntoEnergy})
 must have $T\left(n\right)=\Omega\left(\sqrt{\ln f(n)}\right)$.
Moreover, for all decoding schemes in which $T\left(n\right)$ is faster
than this optimal, $E\ge\Omega\left(\frac{n\ln f(n)}{T\left(n\right)}\right)$.\end{thm}
\begin{IEEEproof}
Suppose that
\begin{equation}
AT=\Theta\left(\sqrt{\ln f(n)}n\right)\label{eq:ATAssumptionForGeneralTheorem}
\end{equation}
Note that from (\ref{eq:ATsquaredTheoremAsFunctionOfFn}), 
\begin{equation}
AT^{2}\ge\Omega\left(n\ln f(n)\right).\label{eq:ATsquaredBoundForThisProof}
\end{equation}

As well, suppose $T\left(n\right)\le O\left(\frac{\sqrt{\ln f(n)}}{g\left(n\right)}\right)$
for some increasing $g\left(n\right)$. Then, from the bound (\ref{eq:ATsquaredBoundForThisProof})
$A\ge\Omega\left(n\sqrt{\ln f(n)}g^{2}\left(n\right)\right)$ (to
prove this, suppose otherwise and derive a contradiction). This implies
then that $AT\ge\Omega\left(\sqrt{\ln f(n)}ng\left(n\right)\right)$,
contradicting (\ref{eq:ATAssumptionForGeneralTheorem}).

Moreover, for all $T\left(n\right)$ growing slower than that required
for optimal energy, this implies that $A\ge\Omega\left(\frac{n\ln\left(f(n)\right)}{T^{2}\left(n\right)}\right)$,
which implies $E\ge\Omega\left(\frac{n\ln f(n)}{T\left(n\right)}\right).$\end{IEEEproof}
\begin{cor}
\label{cor:All-polynomially-low-error}All polynomially-low error
decoding schemes have energy that scales at least as 
\begin{equation}
E\ge\Omega\left(n\sqrt{\ln n}\right).\label{eq:EnergyOfPolynomiallyLower}
\end{equation}
If this optimal is reached, then $T\left(n\right)\ge \Omega (\sqrt{\ln n})$.\end{cor}
\begin{IEEEproof}
This energy lower bound flows from letting $f(n)=\frac{1}{n^{k}}$ and then substituting
this value into (\ref{eq:energyLowerBound}). The time lower bound flows from directly applying Theorem~\ref{thm:generalTheoremForoptimalT}.
\end{IEEEproof}

\subsection{Serial Decoding Scheme Scaling Rules}

Let the number of output nodes in a particular decoder be denoted
$j$ (in a decoding scheme this will be a function of $n$). 
\begin{defn}
A \emph{serial }decoding scheme is one in which $j$ is constant.
\end{defn}
In \cite{blakeConstantPePaper} we considered the case of allowing
the number of output nodes $j$ to increase with increasing block
length. We required an assumption that such a scheme be output regular,
which we define below.
\begin{defn} \cite{blakeConstantPePaper}
An \emph{output regular }circuit\emph{ }is one in which each output
node of the circuit outputs exactly one bit of the computation at
specified clock cycles. This definition excludes circuits where some
output nodes output a bit during some clock cycle and other output
nodes do not during this clock cycle. An \emph{output regular decoding
scheme }is one in which each decoder in the scheme is an output regular
circuit.\end{defn}
\begin{thm}
\label{thm:All-serial--decoding}All serial $f(n)$-decoding schemes
have energy that scales as $\Omega\left(n\ln f(n)\right)$.\end{thm}
\begin{IEEEproof}
The $\Omega\left(n\ln f(n)\right)$ lower bound flows from following
the arguments of the proof of Theorem~2 in \cite{blakeConstantPePaper},
by showing that any decoding scheme in which the area scales less
than $O(\ln f(n))$ cannot be an $f\left(n\right)$-decoding scheme.\end{IEEEproof}
\begin{thm}
\label{thm:All-output-regular}All output regular increasing-output
node $f\left(n\right)$-decoding schemes have energy that scales as
$\Omega\left(n\left(\ln f(n)\right)^{\frac{1}{5}}\right)$.\end{thm}
\begin{IEEEproof}
From the derivations preceding equation (13) in \cite{blakeConstantPePaper},
following a similar argument as in this paper, we divide the circuit
into $M=\Theta\left(\frac{n}{A}\right)$ epochs as before, and divide
the subcircuits into $N=\Theta\left(\frac{A}{\ln f\left(n\right)}\right)$
subcircuits through nested bisections. With this choice, we can follow
the same arguments used in Theorem~3 in \cite{blakeConstantPePaper},
and derive that all $f(n)$-decoding schemes must have 
\[
AT\ge\Omega\left(n\left(\ln\left(f(n)\right)\right)^{\frac{1}{5}}\right).
\]

\end{IEEEproof}

\section{\label{sec:Information-Friction-Section}Information Friction in
Three-Dimensional Circuits}

The ``information friction'' computational energy model was introduced
by Grover in \cite{GroverInfoFriction} and further studied by Vyavahare
\emph{et al.} in \cite{VyavahareBitMeters} and Li \emph{et al.} in
\cite{LiBakshiGroverCompressedSensing}. We generalize (and slightly
modify) this model to three dimensions and use a similar approach
to Grover to obtain some non-trivial lower bounds on the energy complexity
of three dimensional bit-meters decoder circuits, in terms of block
length and probability of error. We will discuss how this approach
can be generalized to models in arbitrary numbers of dimensions. We
present the model below and then prove our main complexity result.
\begin{itemize}
\item A circuit is a grid of computational nodes at locations in the set
$\mathbb{Z}^{3}$, where $\mathbb{Z}$ is the set of integers. Some nodes are\emph{
inputs nodes}, some are \emph{output nodes}, and some are \emph{helper
nodes}. Note that Grover \cite{GroverInfoFriction} considers this
model in terms of a parameter characterizing the distance between
the nodes, but since we are concerned with scaling rules, we will
assume that they are placed at integer locations, allowing us to avoid
unnecessary notation. The Grover paper considered scaling rules in
which nodes are placed on a plane, in which the number of dimensions
$d=2$. In our results we will discuss the case of $d=3$ and afterwards
discuss how the approach can be generalized to an arbitrary
number of spatial dimensions.
\item A circuit is to compute a function of $n$ binary inputs and $k$
binary outputs.
\item At the beginning of a computation, the $n$ inputs to the computation
are injected into the input nodes. At the end of the computation the
$k$ outputs should appear at an output node. A node can be both input
and output.
\item  A node can communicate
messages along its links to any other node, and can receive bits communicated
to them from any other node. 
\item Each node has constant memory, and can compute any computable function
of all the inputs it has received throughout the computation that
is stored in their memory, to produce a message that it can send to
any other node. 
\item We associate a computation with a directed multi-graph, that is, a
set of edges linking the nodes. For every computation, there is one
edge per bit communicated along a link in the computation's associated
multi-graph. The ``cost'' of an edge in such a multi-graph is the
Euclidean distance between the two nodes that it connects. Note that
if a node communicates $m$ bits to another node in a computation,
then that computation's associated multi-graph must have $m$ edges
connecting the two nodes. This multi-graph is called a computation's
\emph{communication multi-graph.}
\item The \emph{energy}, or the \emph{bit-meters}, denoted $\beta$ of a computation is the
sum of the costs of all the edges in the computation's associated
multi-graph (that is, the sum of the Euclidean distances of all the
edges). 
\end{itemize}

\begin{figure} \centering 

\includegraphics[width=3 in]{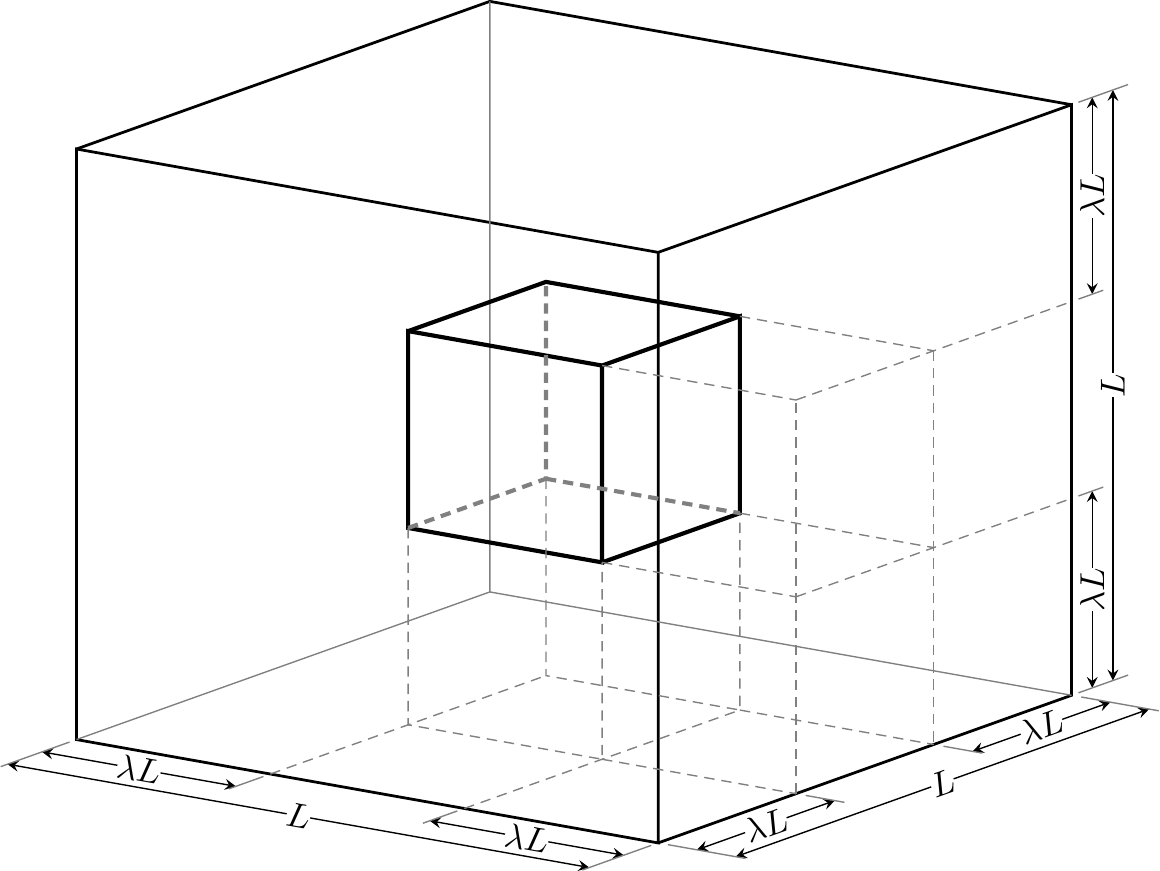}

\caption{A diagram of one nested cube in an $(L, \lambda)$-nested cube grid, with the edge lengths labeled. A nested cube grid is an infinite grid of such nested cubes. The outer cubes each have side length $L$ and the inner cubes each have side length $L(1-2\lambda)$ at a distance $L\lambda$ from the faces of the outer cube.}
\label{fig:nestedCubeGrid}  
\end{figure}

We consider a grid of three-dimensional cubes, with ``inner cubes''
nested within them. This object is a generalization of the ``stencil''
object defined by \cite{GroverInfoFriction}.
\begin{defn}
An \emph{$\left(L,\lambda\right)-$nested cube grid} is an infinite
grid of cubes, with side length $L$ and inner cube side length $L\left(1-2\lambda\right)$.
Note that the inner cubes are centered within the outer cubes. Fig.
\ref{fig:nestedCubeGrid} shows a diagram of one cube in a \emph{$\left(L,\lambda\right)-$nested
cube grid}, to which the reader can refer to visualize this nested
cube structure. A set of nested cube grid parameters is \emph{valid
}if $L>0$ and $0<\lambda<\frac{1}{2}$. 
\end{defn}

Note that a nested cube grid can be placed conceptually on top of
a bit meters circuit. We will consider placing a nested cube grid
in parallel with the Cartesian $3$-space that defines our circuit.
We can specify the position of a nested cube grid that is parallel
to a set of Cartesian coordinates by calling one of the corners of
an outer cube the \emph{origin,} and then specify the location of
its origin. A particular set of parameters for a nested cube grid
and a location for its origin (called its \emph{orientation}) induces
a set of subcircuits, defined below.
\begin{defn}
A \emph{subcircuit}, associated with a particular orientation of a
nested cube grid, is the part of a bit-meters circuit within a particular
outer cube.
\end{defn}
Nodes in any subcircuit can thus be considered to be either inside
an inner cube or outside an inner cube. For any circuit with finite
number of nodes there will thus be some cubes that contain computational
nodes, and some that do not. We can label the subcircuits that contain
nodes with the index $i$. The number of input nodes in cube $i$
we denote $n_{i}$. The number of output nodes in subcircuit $i$
we denote $k_{i}$. Furthermore, we denote the number of input nodes
within the inner cube of subcircuit $i$ as $k_{\mathrm{in},i}$. 
\begin{defn}
We define $k_{\mathrm{in}}=\sum k_{\mathrm{in},i}$, which is the
\emph{the number of output nodes within inner cubes}, which we will
often simply refer to with the symbol $k_{\mathrm{in}}$.
\end{defn}
We will show in Lemma~\ref{lem:3DbitmetersLemma} that there exists
a nested cube grid orientation in which $k_{\mathrm{in}}$ is high.
\begin{defn}
The\emph{ internal bit meters} of a subcircuit $i$ is the length
of all the communication multigraph edges completely within subcircuit
$i$, plus the length of the parts of the edges within subcircuit
$i$. This quantity is denoted with the symbol $\beta_{i}$. Note
that $\beta=\sum_{\mathrm{all\ subcircuits\ j}}\beta_{j}$ (where
we may have to sum over some subcircuits that do not contain any nodes).
\end{defn}
Since a computation has associated with it its communication multi-graph,
for a given subcircuit we can consider the subgraph formed by all
the paths that start outside of the cube and end inside the inner
cube. We can group all the vertices of this graph that start outside
the outer cube and call this the source, and group all vertices inside
an inner cube and call it the sink. For this graph we can consider
its min-cut, the minimum set of edges that, once removed, disconnects
the source from the sink.
\begin{defn}
The\emph{ number of bits communicated from outside a cube to within
an inner cube, }or, \emph{bits communicated,} is the size of this
minimum cut.  For a particular subcircuit $i$ we refer to this quantity
with the symbol $b_{i}$.\end{defn}
\begin{rem}
This quantity is analogous (but not the same) as the quantity $b_{i}$
for the Thompson circuit model from Definition~\ref{def:biBitsCommunicatedToiThSubcircuit},
and thus we use the same symbol. The reader should not confuse these
symbols; the Thompson model definition applies to discussions in Section
\ref{sec:ThompsonModelSection}, and the bit-meters model definition
applies in this section, Section \ref{sec:Information-Friction-Section}. 
\end{rem}
If the $n_{i}$ internal bits of a subcircuit are fixed, then the
subcircuit inside an inner cube will compute a function of the messages
passed from outside the outer cube. Clearly, the size of the set of
possible messages injected into this internal cube is $2^{b_{i}}$
(since $b_{i}$ is the min cut of the paths leading from outside to
inside.)
\begin{lem}
\label{lem:bitMetersProportionalToCommunicatedFromInsideToOut} All
subcircuits with bits communicated $b_{i}$ have internal bit meters
at least $b_{i}\lambda L$.\end{lem}

\begin{IEEEproof}
This result flows from Menger's Theorem \cite{mengersTheorem,GoringshortMengersProof},
which states that any network with min-cut $b_{i}$ has at least $b_{i}$
disjoint paths from source to sink. Each of these paths must have
length at least $\lambda L$ from the triangle inequality.
\end{IEEEproof}
\begin{rem}
	This lemma makes rigorous the idea that to communicate
	$b_{i}$ bits from outside a subcircuit to within its inner square,
	the bit-meters this takes is proportional to the distance from outside
	an outer square to within an inner square ($\lambda L$) and the number
	of bits communicated. \end{rem}

In the lemma below we show that there exists an orientation of any
nested cube grid such that $k_{\mathrm{in}}$ is high.
\begin{lem}
\label{lem:3DbitmetersLemma}For all three dimensional bit-meters
circuits with $k$ output nodes, all valid nested cube grid parameters
$L$ and $\lambda$, there exists an orientation of an $\left(L,\lambda\right)$-nested
cube grid in which the number output nodes within inner cubes ($k_{\mathrm{in}}$)
is bounded by:
\[
k_{\mathrm{in}}\ge\left(1-2\lambda\right)^{3}k
\]
\end{lem}
\begin{rem}
Note that the relative volume of the inner cubes is $\left(1-2\lambda\right)^{3}.$
This lemma says there exists an orientation of any nested cube grid
in which the fraction of output nodes within inner cubes is at least
this fraction, so this result is not surprising.\end{rem}
\begin{IEEEproof}
This is a natural generalization of the Grover result (See Lemma 2
of \cite{GroverInfoFriction}), which uses the probabilistic method.
We consider placing the origin of an $\left(L,\lambda\right)$-nested
cube grid uniformly randomly within a cube of side length $L$ centered
at the origin in the Cartesian $3$-space. We index the $k$ output
nodes by $i$. Let $1_{\mathrm{in,}i}$ be the indicator random variable
that is equal to $1$ if output node $i$ is within an inner cube.
Then, given the uniform measure on the position of the cube, the quantity
$k_{\mathrm{in}}$ is a random variable. We observe:
\begin{eqnarray}
k_{\mathrm{in}} & = & \sum_{i=1}^{k}1_{\mathrm{in,}i}\mathrm{,\ thus}\nonumber \\
E\left(k_{\mathrm{in}}\right) & = & E\left(\sum_{i=1}^{k}1_{\mathrm{in,}i}\right)\nonumber \\
 & = & \sum_{i=1}^{k}E\left(1_{\mathrm{in,}i}\right)\nonumber \\
 & = & \sum_{i=1}^{k}\left(1-2\lambda\right)^{3}\label{eq:averageNumberInside}\\
 & = & k\left(1-2\lambda\right)^{3}\nonumber 
\end{eqnarray}
where in (\ref{eq:averageNumberInside}) we use the observation that,
for each output node, the probability that it is in an inner square
is proportional to the relative area of the inner square. Thus, the
expected value of $k_{\mathrm{in}}$ is $k\left(1-2\lambda\right)^{3}$
and so there must be at least one nested cube grid orientation in
which $k_{\mathrm{in}}$ is greater than or equal to that value.\end{IEEEproof}
\begin{lem}
\label{lem:maxNumberOfniLemma}For all valid nested cube parameters
$L$ and $\lambda$, $n_{i}\le\left(L+1\right)^{3}$ and thus for
sufficiently large $L$ $n_{i}\le2L^{3}$.\end{lem}
\begin{IEEEproof}
Intuitively, there cannot be more than on the order of $L^{3}$ inner
nodes in a cube of volume $L^{3}$. The $\left(L+1\right)^{3}$ bound
comes from considering the corner case of a cube whose sides exactly
touch output nodes.
\end{IEEEproof}

We can now state the main results of this section.
\begin{thm} \label{3DTheorem}
All 3D-bit-meters decoders for a binary erasure channel with erasure
probability $\epsilon$ of sufficiently large block length with block
error probability $P_{e}$ have bit-meters $\beta$ bounded by:
\end{thm}
\[
\beta>\frac{27}{512}\left(\frac{\ln\left(4P_{\mathrm{e}}\right)}{2\ln(\epsilon)}\right)^{\frac{1}{3}}k.
\]

\begin{IEEEproof}
We consider the number of bits communicated from outside a subcircuit
$i$ to within the inner cube of subcircuit $i$ (\textbf{$b_{i}$}).
It must at least be $k_{\mathrm{in},i}$ to overcome the case that
all the input nodes in the entire cube are erased. If this does not
happen, then one of the output nodes must guess at least one bit,
making an error with probability at least $\frac{1}{2}$, formally
justified by Lemma~\ref{lem:GuessingLemma}. This allows us to argue
that:
\begin{eqnarray}
P_{\mathrm{e}} & \ge & P\left(\mathrm{error}|\mathrm{all\ }n_{i}\mathrm{\ output\ bits\ are\ erased}\right)\nonumber \\
 &  & P\left(\mathrm{all\ }n_{i}\mathrm{\ output\ bits\ are\ erased}\right)\nonumber \\
 & \ge & \frac{1}{2}\epsilon^{n_{i}}.\label{eq:probaOfErrorBound}
\end{eqnarray}
If $\beta<\lambda Lk_{\mathrm{in}}$ then there exists a subcircuit
indexed by $i$ in which $b_{i}<k_{\mathrm{in},i}$. Suppose otherwise,
i.e. that $b_{i}\ge k_{\mathrm{in},i}$ for all $i$, then: 
\[
\beta\ge\sum_{\mathrm{all\ subcircuits\ }i}\lambda Lb_{i}=\lambda L\sum b_{i}\ge\lambda L\sum k_{\mathrm{in},i}=\lambda Lk_{\mathrm{in}}
\]
where we apply Lemma~\ref{lem:bitMetersProportionalToCommunicatedFromInsideToOut}
after the first inequality, and for convenience suppress the subscript
on the summation sign after the first instance. This contradicts our
assumption that $\beta<\lambda Lk_{\mathrm{in}}$.

We choose the parameter $L$ in terms of probability of error in order
to derive a contradiction if a circuit does not have high enough bit-meters.
Specifically, we choose 
\begin{equation}
L=\left(\frac{\ln\left(4P_{\mathrm{e}}\right)}{2\ln(\epsilon)}\right)^{\frac{1}{3}}.\label{eq:LintermsofPe}
\end{equation}

Consider the nested cube structure that has $k_{\mathrm{in}}\ge\left(1-2\lambda\right)^{3}k$
that must exist by Lemma~\ref{lem:3DbitmetersLemma}. If $\beta\le\lambda Lk_{\mathrm{in}}$
then there must exist a subcircuit $i$ that has less than $k_{\mathrm{in},i}$
bits injected into it from outside the subcircuit to within its inner
cube. Thus: 
\[
\mathrm{if\ }\beta\le\lambda Lk_{\mathrm{in}}\mathrm{\ then\ }P_{\mathrm{e}}\overset{(a)}{\ge}\frac{1}{2}\epsilon^{n_{i}}\overset{(b)}{\ge}\frac{1}{2}\epsilon^{2L^{3}}\overset{(c)}{\ge}2P_{\mathrm{e}}
\]
where (a) flows from (\ref{eq:probaOfErrorBound}), (b) from Lemma~\ref{lem:maxNumberOfniLemma}, and (c) from the evaluation of this
expression by substituting (\ref{eq:LintermsofPe}). This is a contradiction.
Thus, all bit meters decoders must have
\begin{eqnarray*}
\beta & > & \lambda Lk_{\mathrm{in}}\\
\beta & > & \lambda\left(1-2\lambda\right)^{3}Lk\\
 & \ge & \lambda\left(1-2\lambda\right)^{3}\left(\frac{\ln\left(4P_{\mathrm{e}}\right)}{2\ln(\epsilon)}\right)^{\frac{1}{3}}k.
\end{eqnarray*}
The second inequality flows from the fact that we are considering
the nested cube structure in which $k_{\mathrm{in}}\ge\left(1-2\lambda\right)^{3}k$
that must exist by Lemma~\ref{lem:3DbitmetersLemma}. We may choose any valid $\lambda$ to maximize this bound, and letting  $\lambda=\frac{1}{8}$ gives us: 
\[
\beta>\frac{27}{512}\left(\frac{\ln\left(4P_{\mathrm{e}}\right)}{2\ln(\epsilon)}\right)^{\frac{1}{3}}k.
\]
\end{IEEEproof}
\begin{rem}
Note that this argument naturally generalizes to $d$-dimensional
space, in which all $d$-dimensional bit-meters decoders have energy
that scales as $\beta\ge\Omega\left(\left(\ln\left(P_{\mathrm{e}}\right)\right)^{\frac{1}{d}}k\right)$.
The key step in the proof to be altered is in a modification of Lemma~\ref{lem:maxNumberOfniLemma} and a choice of $L=c\left(\frac{\ln\left(4P_{\mathrm{e}}\right)}{\ln(\epsilon)}\right)^{\frac{1}{d}}$in
line \ref{eq:LintermsofPe} of the proof for some constant $c$ that
may vary depending on the dimension. This implies, among other things,
that exponentially low probability of error decoding schemes implemented
in $d$-dimensions have bit-meters energy that scales as $\Omega\left(n^{1+\frac{1}{d}}\right)$.
Obviously, the most engineering-relevant number of dimensions
$d$ for this type of analysis are $d=2$ and $d=3$.
\end{rem}

\section{Encoder Lower Bounds\label{sec:encoderLowerBounds}}

In terms of scaling rules, all the decoder lower bounds presented
herein can be extended to encoder lower bounds. The main structure
of the decoder lower bounds (inspired by \cite{groverFundamental, GroverInfoFriction}) involves dividing
the circuit into a certain number of subcircuits. Then, we argue that
if the bits communicated within the circuit is lower, then there must
be one subcircuit where the bits communicated to it are less than
the bits it is responsible for decoding. If all the inputs bits in
that circuit are erased, the decoder must make an error with probability
at least $1/2$.

In the encoder case, we also take inspiration from \cite{groverFundamental, GroverInfoFriction}. In this
case, the $n$ outputs of the encoder circuit can be divided into
a certain number of subcircuits. Then we consider the bits communicated
\emph{out} of each subcircuit. This quantity must be proportional
to the number of output bits in each subcircuit. Otherwise, there
will be at least one subcircuit where the number of bits communicated
out is less than the number of output nodes in the subcircuit. Call
these bits that were not fully communicated out of this subcircuit
\emph{$Q$.} Suppose that once the output bits of the encoder are
injected into the channel, all the bits in $Q$ are erased. Now, the
decoder must use the other bits of the code to decode. But, the subcircuit
containing $Q$ in the encoder communicated less than $\left|Q\right|$
bits to the other outputs of the encoder. By directly applying Lemma~\ref{lem:GuessingLemma}, we see that no matter what function the decoder computes, it
must make an error with probability at least $1/2$.
An argument of this structure and following exactly the structure of Theorems \ref{thm:TheoremRelatingfofntoEnergy}, \ref{thm:generalTheoremForoptimalT}, \ref{thm:All-serial--decoding}, \ref{thm:All-output-regular}, and \ref{3DTheorem}  for the decoders gives us the following theorems, whose proofs are omitted.

\begin{thm}
All fully-parallel $f(n)$-encoding schemes with number of
clock cycles $T(n)$ have energy
\[
E(n)\ge\Omega\left(\frac{n\log(f(n)}{T(n)}\right)
\]
with optimal lower bound of $E\ge\Omega\left(n\sqrt{\log f(n)}\right)$
when $T(n)\ge\sqrt{\log(f(n))}$.

All serial, $f(n)$-encoding schemes have energy that scales as
\[
E(n)\ge\Omega\left(n\log f(n)\right).
\]
All increasing output node, output-regular $f(n)$-encoding schemes
have energy that scales as 
\[
E(n)\ge\Omega\left(n\log^{1/5}\left(f(n)\right)\right).
\]
Finally, all three-dimensional, bit-meters encoding schemes associated
with block error probability $P_{\mathrm{e}}$ have energy that scales 

\[
E(n)\ge\Omega(n\ln P_{\mathrm{e}}).
\]
\end{thm}

\section{Limitations of Results\label{sec:Limitations-of-Result}}

There are a number of weaknesses in the models we have used. Firstly,
our results are asymptotic. 
For some set block error probability and rate, there may be a specific
circuit that reaches this block error probability using a circuit
design methodology that does not generalize to scale in a way as predicted
by our theorems. 

Note that our quantity $T$ refers to number of clock cycles, which
reflects one of the main ``time costs'' in a circuit computation. In real circuits, the ``time cost'' of a computation involves two
parameters: the number of clock cycles required, and the time it takes
to do each clock cycle. In our model, we do not consider the time
per clock cycle. In real circuits, this quantity often varies
with wire lengths. We do not consider this in our model.

A particular weakness of the Thompson model we use is that it does
not consider a quantity called \emph{switching activity factor}. In
circuit design, this quantity is the fraction of the circuit that
``switches'' during the course of the computation. And
yet, our model assumes a switching activity factor of $1$. Thus, in terms
of scaling rules, the Thompson model should be considered applicable
only to computational schemes in which the switching activity factor
does not change with increasing input sizes. On the other hand, the
information-friction model accounts for the possibility of schemes
in which switching activity factor changes with increasing block length,
so, combined with the results of Grover, \cite{GroverInfoFriction},
the asymptotic energy lower bounds we derive apply.

\section{\label{sec:Other-Energy-Models}Other Energy Models of Computation}

There has been some work on energy models of computation different
from the Thompson energy models and Grover information friction models,
and herein we provide a short review.

In \cite{binghamGreenstreet2012}, Bingham \emph{et al. }classify
the tradeoffs between the ``energy'' complexity of parallel algorithms
and ``time'' complexity for the problem of sorting, addition, and
multiplication using a model similar to, but not the same as the model we use. In the
grid model used by these authors, a circuit is composed of processing elements laid
out on a grid, in which each element can perform an operation. In this model the circuit designer has choice over the speed of each
operation, but this comes at an energy cost. Real circuits run at
higher voltages can result in lower delay for each processing element but higher energy
\cite{Honeisenmead}. The model used by the authors in \cite{binghamGreenstreet2012}
captures some of this fundamental tradeoff. Note that our model
assumes constant voltage. Non-trivial
results that show how real energy gains can occur by lowering voltages
in decoder circuits have been studied in \cite{Primeau}, but we do
not study this here.

Another energy model of computation was presented by Jain \emph{et al.} in \cite{jainMolnarModel}. This model introduced an augmented Turing
machine, a generalization of the traditional Turing machine \cite{turing1936}. The authors introduce a transition
function, mapping the current instruction being
read, the current state, the next state and the next instruction to
the ``energy'' required to make this transition.
This model (once the transition function is clearly defined for a
specific processor architecture) would be good for the algorithm designer
at the software level. However, we do not believe this model informs
the specialized circuit designer. The Thompson model which we analyze,
on the other hand, can include, as a special case, the energy complexity
of algorithms implemented on a processor, as our model allows for
a composition of logic gates to form a processor.

 Landauer \cite{Landauer1961}  derives that the energy required to erase one bit of information is at least $kT\ln 2$, where $k$ is Boltzmann's constant, and $T$ is the temperature. Thus, a fundamental limit of computation comes from having to erase information. Of course, it may be possible to do reversible computation in which no information is erased that can use arbitrarily small amounts of energy, but such circuits must be run arbitrarily slowly. This suggests a fundamental time-energy tradeoff different from the tradeoff discussed herein. Landauer \cite{LandauerReview}, Bennett \cite{bennetReview1982} and Lloyd \cite{sethLloydNature} provide detailed discussions and bibliographies on this line of work. Demaine \emph{et al.} \cite{DemaineLandauerEnergy2016} extract a mathematical model from this line of work and analyze the energy complexity of various algorithms within this model. Note that the Thompson model we use is one informed by how modern VLSI circuits are created, even though they operate at energies far above ultimate physical limits.

\section{Future Work\label{sec:Future-Work}}

Currently, our work on lower bounds has not be extended to other channels, like the additive white Gaussian noise channel. Perhaps more interesting, however, is the question, do there exist
polynomially low probability of error decoding schemes with energy
that closely matches (\ref{eq:EnergyOfPolynomiallyLower}) of Corollary~\ref{cor:All-polynomially-low-error}, i.e., one with energy that
scales as $\Omega\left(n\sqrt{\ln n}\right)$? This may have significantly
lower energy than an exponentially-low error decoding scheme, and
may provide sufficient error control performance. We do not know whether
such a decoding scheme exists and this remains an important open question.
It may be that decoding strategies with energy that scales like this
are already invented but have simply not been analyzed in terms of
their energy complexity.

The decoding problem for communication systems is a special case of
the more general problem of inference. Well known algorithms used
for inference, for example the Sum-Product Algorithm \cite{KschischangFactorGraphs}
and variational methods \cite{wainwrightVariationalInference}, include
Gallager's low-density parity-check decoding algorithms as a special
case \cite{GallagerLDPC}. Thus, we conjecture that there may be similar
tradeoffs between energy, latency, and reliability  in circuits that
perform inference.

\bibliographystyle{IEEEtran}
\bibliography{bibtextDoc}
\end{document}